\documentclass[12pt]{article}
\usepackage{amsfonts,amsmath,amssymb,graphicx}

\title{Two Arrows of Time in Nonlocal Particle Dynamics}
\author{
  Roderich Tumulka\footnote{Department of Mathematics, Rutgers University, 
	110 Frelinghuysen Road, Piscataway, NJ 08854-8019, USA. 
	E-mail: tumulka@math.rutgers.edu}
}
\date{July 21, 2007}

\newcommand{\CCC}{\mathbb{C}} 
\newcommand{\I}{\mathrm{i}} 
\newcommand{\1}{\mathbf{1}} 
\newcommand{\const}{\mathrm{const}} %
\newcommand{\st}{(\mbox{space-time})}

\begin{document}
\maketitle

\begin{abstract}
Considering what the world would be like if backwards causation were
possible is usually mind-bending.  Here I discuss something that is
easier to study: a toy model that incorporates a very restricted sort of
backwards causation. It defines particle world lines by means of a kind of 
differential delay equation with negative delay. The model presumably prohibits 
signalling to the past and superluminal signalling, but allows nonlocality 
while being fully covariant.  And that is
what constitutes the model's value: it is an explicit example of the possibility of
Lorentz invariant nonlocality. That is surprising in so far as many authors
thought that nonlocality, in particular nonlocal laws for particle world lines, 
must conflict with relativity. 
The development of this model was inspired by the search for a
fully covariant version of Bohmian mechanics.
\end{abstract}


In this paper I will introduce to you a dynamical system---a law of
motion for point particles---that has been invented \cite{arrow} as a
toy model based on Bohmian mechanics. Bohmian mechanics is a version
of quantum mechanics with particle trajectories; see \cite{Gol01} for
an introduction and overview. What makes this toy model remarkable is
that it has two arrows of time, and that precisely its having two
arrows of time is what allows it to perform what it was designed for: to have
effects travel faster than light from their causes (in short,
\emph{nonlocality}) without breaking Lorentz invariance. Why should
anyone desire such a behavior of a dynamical system? Because Bell's
nonlocality theorem \cite{Bell} teaches us that any dynamical system
violating Bell's inequality must be nonlocal in this sense. And Bell's
inequality is, after all, violated in nature. 

It is easy to come up with a nonlocal theory if one assumes that
one of the Lorentz frames is preferred to the others: simply assume a
mechanism of cause and effect (an interaction in the widest
sense) that operates \emph{instantaneously} in the preferred
frame. That is what nonrelativistic theories usually do. In other
frames, these nonlocal effects will either travel at a superluminal
($>c$) but finite velocity or precede their causes by a short time
span. Of course, causal loops can't arise since in the
preferred frame effects never precede causes; yet the entire notion of
a preferred frame is against the spirit of relativity.
Without a preferred frame, to find a nonlocal law of motion is tricky,
and much agonizing has been spent on this. About one way to achieve
this you will learn below.

Let's come back first to the two arrows of time. They are opposite
arrows, in fact. But unlike the arrows considered in Lawrence
Schulman's contribution to this volume, they are not both
thermodynamic arrows. One of the two is the thermodynamic
arrow. Let's call it $\Theta$. It arises, as emphasized first by
Ludwig Boltzmann and in this conference by Schulman, not from whichever
asymmetry in the microscopic laws of motion, but from boundary
conditions. That is, from the condition that the initial state of the
universe be taken from a particular subset of phase space
(corresponding to, say, a certain low entropy macrostate), while the
final state is not subjected to any such conditions---except in some
scenarios studied by Schulman. The dynamical laws considered in
discussions of the thermodynamic arrow of time are usually time
reversal invariant. But not so ours! It explicitly breaks time
symmetry, and that is how another arrow of time comes in: an arrow of
microscopic time asymmetry, let's call it $C$. Such an arrow must be
assumed before writing down the equation of motion, which will be equation
(\ref{eqmotion}) below. In addition, the equation of motion is easier to
solve in the direction $C$ than in the other direction. Doesn't it
seem ugly and unnatural to introduce a time asymmetry? Sure, but we
will see it buys us something: Lorentz invariant nonlocality.

Recall that such an arrow is simply absent in Newtonian mechanics and other
time symmetric theories. So it is not surprising that the microscopic
arrow $C$ is not the source of the macroscopic time arrow $\Theta$,
even more, the direction of $\Theta$ is completely independent of the
direction of $C$. $\Theta$ depends on boundary conditions, and not on
the details of the microscopic law of motion. In our case,
$\Theta$ will indeed be opposite to $C$. Since inhabitants of a
hypothetical universe will regard the thermodynamic arrow as their
natural time arrow, related to macroscopic causation, to memory, and
to apparent free will, you should always think of $\Theta$ as pointing
towards the future, whereas $C$ is pointing to what we call the past.

It's time to say what the equation of motion is. The equation is
intended to be as close to Bohmian mechanics as possible, to be an
immediate generalization, and to have Bohmian mechanics as its
nonrelativistic limit. To remind you of how Bohmian mechanics works,
you take the wave function (which is supposed to evolve according to Schr\"odinger's
equation---without ever having to collapse), plug in the positions of
all the particles (here is where a notion of simultaneity comes in),
and from that you compute the velocity of any particle by applying a
certain formula, Bohm's law of motion, which amounts to dividing the
probability current by the probability density. Now, for a
Lorentz-invariant version, we first have to worry about the wave
function.

There are three respects in which the wave function of nonrelativistic
quantum mechanics (or Bohmian mechanics, for that matter) conflicts
with relativity: (a)~the dispersion relation $E=p^2/2m$ at the basis
of the Schr\"odinger equation is nonrelativistic, (b)~the wave
function is a function of $3N$ position coordinates but only one time coordinate,
(c)~the collapse of the wave function is supposed
instantaneous. While (a) has long been solved by means
of the Klein--Gordon or Dirac equation, it is too early
for enthusiasm since we still face (b) and (c). We will worry about
(c) later, and focus on (b) now. The obvious answer is to introduce a
wave function $\psi$ of $4N$ coordinates, that is one time coordinate
for each particle, in other words $\psi$ is a function on
$\st^N$. You get back the nonrelativistic function of $3N+1$
coordinates after picking a frame and setting all time coordinates
equal. Such multi-time wave functions were first considered by Dirac
\textit{et al.}\ in 1932 \cite{Dirac}, but what they didn't mention
was that the $N$ time evolution equations
\begin{equation}\label{multitime}
  \I\hbar \frac{\partial \psi}{\partial t_i} = H_i \psi \quad
  \mbox{for } i\in\{1,\ldots,N\}
\end{equation}
needed for determining $\psi$ from initial data at $t=0$ do not always
possess solutions. They are usually inconsistent. They are only
consistent if the following condition is satisfied:
\begin{equation}\label{commutator}
  [H_i, H_j]=0 \quad \mbox{for }i\neq j\,.
\end{equation}
This is easy to achieve for non-interacting particles and tricky in
the presence of interaction. Indeed, to my knowledge it has never been
attempted to write down consistent multi-time equations for many
interacting particles, although this would seem an obvious and
highly relevant problem if one desires a a manifestly covariant
formulation of relativistic quantum mechanics. We will here, however,
stay on the easy side and simply consider a system of non-interacting
particles. We take the multi-time equations to be Dirac equations in
an external field $A_\mu$,
\begin{equation}\label{Diraceq}
  \1\otimes \cdots \otimes \!\! \underbrace{\gamma^\mu}_{i\mathrm{th}\,\,
  \mathrm{place}} \!\! \otimes \cdots \otimes \1 \, \Bigl(
  \I\frac{\partial}{\partial x_i^{\mu}}-eA_{\mu}(x_i) \Bigr)\, \psi =
  m\psi
\end{equation}
where $\psi:\st^N \to (\CCC^4)^{\otimes N}$, and $e$
and $m$ are charge and mass, respectively. The corresponding
Hamiltonians commute trivially since the derivatives act on different
coordinates and the matrices on different indices.

Such a multi-time Dirac wave function naturally defines a tensor field 
\begin{equation}\label{Jdef}
  J^{\mu_1\ldots \mu_N} := \overline\psi \, \gamma^{\mu_1} \otimes
  \cdots \otimes \gamma^{\mu_N} \, \psi\,,
\end{equation}
and according to the original Bohmian law of motion (for Dirac wave
functions), the 4-velocity of particle $i$ is, in the preferred frame,
\begin{equation}\label{Bohm}
  \frac{dQ_i^\mu}{ds} \propto J^{0\ldots \stackrel{i}{\mu} \ldots 0}(Q_1, \ldots, Q_N)
\end{equation}
where only the $i$th index of $J$ is nonzero, and $Q_i^\mu(s)$ is the world line parameterized by proper time, or indeed
by any other parameter since a law of motion need only (and \eqref{Bohm} does only)
specify the \emph{direction} in space-time of the tangent to the world line. 
The coordinates taken for the other particles are their
positions \emph{at the same time}, $Q_j^0 = Q_i^0$. Instead of a
Lorentz frame, one can take any foliation of space-time into spacelike
hypersurfaces for the purpose of defining simultaneity-at-a-distance
\cite{HBD}. The theory I'm about to describe, in contrast, uses the
hypersurfaces naturally given by the Lorentzian structure on
space-time: the light cones. More precisely: the future light
cones---and that is how the time asymmetry comes in.

\begin{figure}[h]
\begin{center}
\includegraphics[width=.5\textwidth]{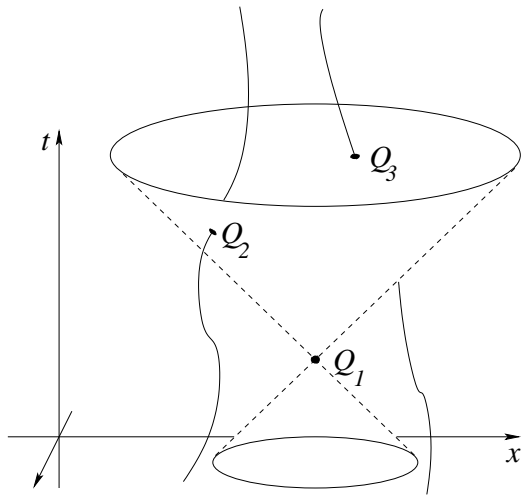}
\end{center}
\caption[]{How to choose the $N$ space-time points where to evaluate
the wave function, as described in the text.}
\label{lightc}\end{figure}

So here are the steps: first solve (\ref{Diraceq}), so you know $\psi$
on $\st^N$. Then, compute the tensor field $J$ on $\st^N$ according to
(\ref{Jdef}). For determining the velocity of particle $i$ at
space-time point $Q_i$, find the points $Q_j$, $j \neq i$, where the other
particles cross the future light cone of $Q_i$, as depicted in
Figure~\ref{lightc}. Plug these $N$ space-time points into the field $J$
and get a single tensor. Find out what the 4-velocities $u_j^{\mu}$
of the other particles at $Q_j$, $j \neq i$, are. Use these to contract all but one
index of $J$. We postulate that the resulting vector is, up to an irrelevant
proportionality factor, the 4-velocity we've been looking for:
\begin{equation}\label{eqmotion}
  \frac{dQ^{\mu_i}_i}{ds} \propto J^{\mu_1 \ldots \mu_N}(Q_1, \ldots, Q_N)\,
  \prod_{j\neq i} u_{j\mu_j}(Q_j)\,.
\end{equation}
One can show \cite{arrow} that this 4-velocity is always timelike or null.

This law of motion is what can be called an ordinary differential
equation with advanced arguments, or a differential delay equation 
with negative delay, because the velocity depends on the
positions (and velocities) of other particles at future times, indeed
with a \emph{variable} delay span $Q_j^0 - Q_i^0$. It may seem to
complicate things considerably that what happens here depends on the
\emph{future} rather than past behavior of the other particles, but
that is an artifact of perspective: look at the equation of motion
(\ref{eqmotion}) in the other time direction, that is in the direction
$C$, and notice it now has only \emph{retarded} arguments. That is a
more familiar sort of differential delay equation that gives rise to
no logical or causal problems. So this theory, although involving a
mechanism of backwards causation, is provably paradox free, since no
causal loops can arise: first solve the wave equation for $\psi$ in
the usual direction $\Theta$, then solve the equation of motion in the
opposite direction $C$.

Unfortunately, there is no obvious probability measure on the set of
solutions to (\ref{eqmotion}). This is different from the situation in
Bohmian mechanics, where the $|\psi|^2$ distribution is conserved, a
fact crucial for the probability predictions of that theory. The lack
of such a measure for the model considered here makes it impossible to
say whether or not this theory violates Bell's inequality, which is a
relation between probabilities. But this law of motion takes what is
perhaps the biggest hurdle on the way towards a fully covariant law of
motion conserving the $|\psi|^2$ distribution, by fulfilling what Bell's theorem
says is a necessary condition: nonlocality.  I should add that in the
nonrelativistic limit, the future light cone approaches the hyperplane
$t=\const.$\ and the law of motion approaches the original Bohmian
law of motion (\ref{Bohm}), conserving $|\psi|^2$.

How does nonlocality come about in this model? That has to do with the
two arrows of time, pointing in opposite directions. Had we chosen
them to point in the same direction, the theory would have been local,
because what happens at $Q_i$ would only depend on (what we call) the
past light cone. But in this model, we evaluate $\psi$ on the future
light cone of $Q_i$, which means $\psi$ has, in its multi-time evolution,
gone through all the external fields at spacelike separation from
$Q_i$.  And that is how the velocity at $Q_i$ may be influenced by the
field imposed by an experimenter at spacelike separation from $Q_i$.

And what is the story then about problem (c) above, the instantaneous
collapse?  The first thing to say is that collapse is not among the
basic rules of this model, or any Bohmian theory.  That simply
disposes of problem (c).  But something more should be said, since the
collapse rule can be derived in Bohmian mechanics: even if the wave
function of Schr\"odinger's cat remains forever a superposition,
\emph{the cat itself} (formed by the particles) is either dead or
alive, with probabilities determined by $|\psi|^2$. Moreover, since the wave packet
of the dead cat (i.e., the corresponding term in the superposition) and that of
the live cat have disjoint supports in configuration space, the wave packet
of the dead cat doesn't influence the motion of the
live cat (nor vice versa).  In the model we are concerned with here, everything just
said still applies, except that the model doesn't define any probabilities.

The model thus shows that a relativistic theory of particle world lines can indeed be nonlocal. Let me also point to another consequence: It has often been claimed that Bell's nonlocality proof excludes relativistic Bohm-type theories. This claim has always been inappropriate because Bell's proof actually shows that \emph{any} serious version of quantum mechanics, Bohm-like or not, must be nonlocal; now we see that the claim is also inappropriate in another way, as nonlocality actually doesn't imply a conflict with relativity. Finally, let me add that a fully covariant version has been developed for a different quantum theory without observers, the GRW theory \cite{rGRWf}. Also this model uses time-asymmetric laws, but not backwards causation.

To this day, thinking about time, time's arrows, and relativity
remains a source of the unexpected.

\medskip

\noindent \textit{Acknowledgement.} I wish to thank Sheldon Goldstein
for his comments on a draft of this paper.

\end{document}